\documentclass[twocolumn,showpacs,prd]{revtex4}
\usepackage{mathrsfs,bm}
\usepackage{longtable,lscape}
\usepackage{txfonts}
\usepackage{amssymb}
\usepackage{indentfirst}
\usepackage{graphicx,,booktabs}
\usepackage{color}
\usepackage{amssymb}

\usepackage[dvipdfm,  
            pdfborder=001,   %
            ]{hyperref}

\begin{document}
\title{New structure around 3250 MeV in the baryonic $B$ decay and the $D^*_0(2400)N$ molecular hadron}
\author{Jun He$^{1,2}$}\email{junhe@impcas.ac.cn}
\author{Dian-Yong Chen$^{1,2}$}\email{chendy@impcas.ac.cn}
\author{Xiang Liu$^{1,3}$\footnote{Corresponding author}}\email{xiangliu@lzu.edu.cn}
\affiliation{
$^1$Research Center for Hadron and CSR Physics,
Lanzhou University and Institute of Modern Physics of CAS, Lanzhou 730000, China\\
$^2$Nuclear Theory Group, Institute of Modern Physics of CAS, Lanzhou 730000, China\\
$^3$School of Physical Science and Technology, Lanzhou University, Lanzhou 730000,  China}
\date{\today}

\begin{abstract}
In this work, we first propose the isovector $nD^{*}_0(2400)^0$ molecular state to explain the
enhancement structure around 3250 MeV ($X_c(3250)^0$) in the $\Sigma_c^{++}\pi^-\pi^-$ invariant mass spectrum newly observed by the BaBar Collaboration. Under this molecular state configuration, both the analysis of the mass spectrum and the study of its dominant decay channel can well depict its resonance parameters measured by BaBar. Our investigation also shows that
the isovector $nD^{*}_0(2400)^0$ molecular state can decay into $\Sigma_c^{++}\pi^-\pi^-$, which is consistent with the experimental observation. These studies provide the direct support to the isovector $nD^{*}_0(2400)^0$ molecular state assignment to $X_c(3250)^0$. 
\end{abstract}


\maketitle

The BaBar Collaboration recently reported a new enhancement structure in the $\Sigma_c^{++}\pi^-\pi^-$ invariant mass spectrum of the $B^-\to \Sigma_c^{++}\bar{p}\pi^-\pi^-$ decay. The mass and width are given by $M=3245\pm20$ MeV and $\Gamma=108\pm 6$ MeV, respectively \cite{babar}. We will
call the new structure as $X_c(3250)^0$ in this work. In terms of the decay channel observed, we conclude that $X_c(3250)^0$ is an isotriplet with charm number $C=+1$. It is an intriguing research topic to understand this new enhancement structure observed in the baryonic $B$ decay.
In this work, we propose a novel approach to explain the BaBar's observation of $X_c(3250)^0$, which
can be naturally explained as a molecular hadron composed of a charmed meson $D_0^*(2400)$ and a nucleon $N$. In the following, we illustrate why the explanation of the $D^*_0(2400)N$ molecular hadron is reasonable for the observed $X_c(3250)^0$ in detail.

A molecular explanation requires that the mass of the observed state should be below and close to the sum of masses of its components. The $X_c(3250)$ newly observed by BaBar just meets this necessary condition because the mass is near the threshold of $D_0^*(2400)$ and $N$.
For the $D^*_0(2400)N$ molecular system, the flavor wave function is written as
\begin{eqnarray}
I&=&1:\,\left\{
  \begin{array}{l}
|X_c(3250)^0\rangle=|D_0^*(2400)^0n\rangle\\
|X_c(3250)^+\rangle=\frac{|D_0^*(2400)^+n\rangle-|D_0^*(2400)^0p\rangle}{\sqrt{2}}\\
|X_c(3250)^{++}\rangle=|D_0^*(2400)^+p\rangle
\end{array} \right. ,\label{t}\\
I&=&0:\,|Y_c(3250)^+\rangle=\frac{|D_0^*(2400)^+n\rangle+|D_0^*(2400)^0p\rangle}{\sqrt{2}},\label{s}
\end{eqnarray}
whose expressions correspond to isotriplet and isosinglet, respectively, where $|X_c(3250)^0\rangle$ is the flavor wave function of $X_c(3250)^0$, etc. Under the assignment of the $D^*_0(2400)N$ molecular state, we further deduce the quantum number of $X_c(3250)$ as $I(J^P)=1(\frac{1}{2}^+)$

With the neutral $X_c(3250)^0$ as an example, we first carry out the calculation of the binding energy for the $D^*_0(2400)N$ molecular system. According to the mass values of the $D^*_0(2400)^0$ and neutron $n$ listed in Particle Data Group (PDG) \cite{Nakamura:2010zzi}, we can obtain the binding energy of $X_c(3250)^0$ as $\sim -13$ MeV, with $M_{D_0^{*}(2400)^0}=2318$ MeV and $M_{n}=940$ MeV. Then we have $M_{D_0^{*}(2400)^0}+M_{n}=3258\, \mathrm{MeV}\,>M_{\Xi_c(3250)}$.
In the following, we will examine whether it is reasonable to explain $X_c(3250)^0$ as the $n D^*_0(2400)^0 $ molecular hadron.
To deduce the effective potential between $D^*_0(2400)^0$ and $n$, we adopt the effective Lagrangian of the light mesons interacting with the charmed meson $D_0^*(2400)$ or nucleon, {\it i.e.},
\begin{eqnarray}
	{\cal L}_{mD_0^*D_0^*}&=& -i\frac{\beta'g_V}{\sqrt{2}}D^*_{0b}D^{*\dag}_{0a}(2iv\cdot\mathbb{V}_{ba})
+2g'_\sigma\sigma D^*_{0a}D^{*\dag}_{0a},	\\
	{\cal L}_{mNN}&=&	 -\sqrt{2}g_{VNN}\bar{N}_b\left(\gamma^\mu+\frac{k\,\sigma^{\mu\nu}}{2m_N}\partial^\nu\right)
\mathbb{V}_{\mu,ba}N_a+g_{\sigma NN}\bar{N}\sigma N\nonumber\\
\end{eqnarray}
with $v=(1,{\bm 0})$, $\sigma^{\mu\nu}=\frac{i}{2}(\gamma^\mu\gamma^\nu
-\gamma^\nu\gamma^\mu)$ and the vector matrix $\mathbb V$
\begin{eqnarray*}
	{\mathbb V}=\left(\begin{array}{ccc}
		\frac{1}{\sqrt{2}}\rho^0+\frac{1}{\sqrt{2}}\omega&\rho^+ \\
		\rho^-&-\frac{1}{\sqrt{2}}\rho^0+\frac{1}{\sqrt{2}}\omega  		
\end{array}\right).\label{v}
\end{eqnarray*}
The index $m$ in ${\cal L}_{mD_0^*D_0^*}$ and ${\cal L}_{mNN}$ denotes the light meson.
The coupling constants involved in this work are given by $\beta'=1$, $g'_\sigma=-0.76$~\cite{Casalbuoni:1996pg,Bardeen:2003kt},
$g^2_{\rho NN}/4\pi=0.84$, $g^2_{\omega NN}/4\pi=20$, {$g_V=m_\rho/f_\pi=5.8$,} $g^2_{\sigma NN}/4\pi=5.69$ and $\kappa=6.1(0)$ for $\rho(\omega)$ \cite{Machleidt:2000ge,Cao:2010km,Tsushima:1998jz,Engel:1996ic}, respectively, where we follow
the convention of the signs of coupling constants in Refs.~\cite{Cao:2010km,Tsushima:1998jz,Engel:1996ic}. The mass of the exchanged sigma is taken as 660 MeV.

\begin{figure}[htb]
\centering
\begin{tabular}{cc}
\scalebox{0.83}{\includegraphics{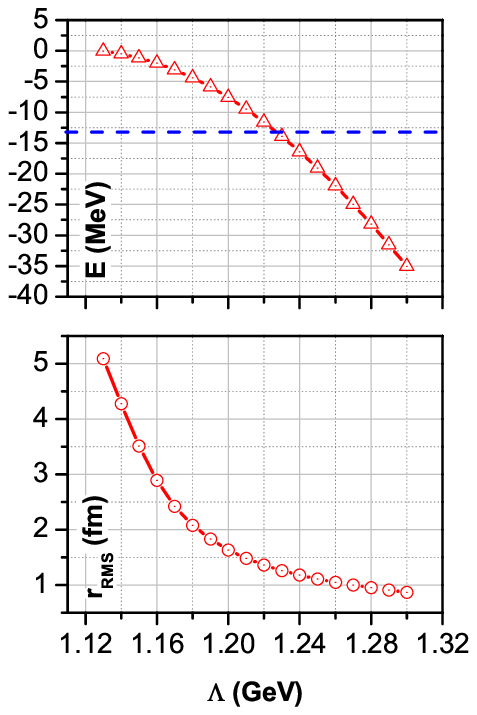}}\raisebox{1ex}{\scalebox{0.83}{\includegraphics{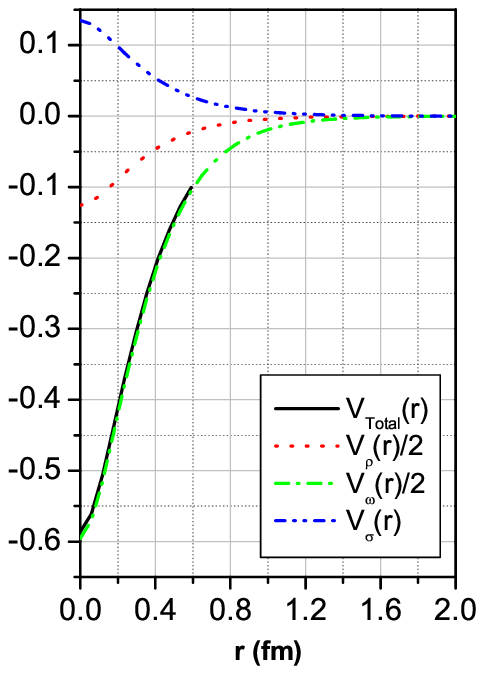}}}
\end{tabular}
\caption{(color online). The obtained bound state solution (binding energy $E$ and root-mean-square $r_{RMS}$) of $X_c(3250)^0$ dependent on cutoff $\Lambda$. Here, the blue dashed line corresponds to $E=-13$ MeV derived by the mass difference $M_{X_c}-M_{D_{0}^*}-M_n$. In addition, we also present the variation of the total effective potential and the subpotentials in Eq. (\ref{total}) in $r$.
\label{Fig: mass}}
\end{figure}

The effective potential of $X_c(3250)^0$ in the coordinate space is given by
\begin{eqnarray}
\mathcal{V}_{\mathrm{Total}}(r)&=&\frac{1}{2}\mathcal{V}_\rho(r)+\frac{1}{2}\mathcal{V}_\omega(r)+\mathcal{V}_\sigma(r) \label{total}
\end{eqnarray}
with
$\mathcal{V}_V(r)=-2\beta'g_{VNN}g_V\, Y(\Lambda,m_V,r)$ and
$\mathcal{V}_\sigma(r)=-2g'_\sigma g_{\sigma NN}\,Y(\Lambda,m_\sigma,r)$,
where the $Y(\Lambda,m,r)$ function is defined as
\begin{eqnarray*}
Y(\Lambda,m,r)=\frac{1}{4\pi r}\bigg[e^{-mr}-e^{-\Lambda r}
-\frac{\Lambda^2-m^2}{2\Lambda}re^{-\Lambda r}\bigg].
\end{eqnarray*}
By solving the Schr\"{o}dinger equation with the effective potential obtained in Eq. (\ref{v}), the dependence on the cutoff $\Lambda$ of the bound state solution for $X_c(3250)^0$
is shown in Fig.~\ref{Fig: mass}. When $\Lambda=1.23$ GeV, the theoretical result of the binding energy for $X_c(3250)^0$ is consistent with the experimental data \cite{babar}. Usually, in the one-boson exchange model the general criteria of forming a molecular state is that we can obtain negative binding energy of this system and the corresponding cutoff $\Lambda$ should be around 1 GeV. Thus, by our calculation, we can conclude that $D_0^*(2400)$ and nucleon can form a loosely bound state with the small binding energy since the adopted $\Lambda$ value is close to 1 GeV.

As shown in Fig. \ref{Fig: mass}, the potential of the $D_0(2400)^0 n$ molecular state is mainly from $\rho$ exchange, where the $\pi$ exchange is absent. It is well known that the vector meson exchanges provide short-range interaction. Usually, the necessary condition of forming a bound state requires that  the mass of the exchanged meson is larger than the widths of the components of this bound state. We notice that the $\rho$ meson mass ($m_\rho=770$ MeV) is larger than the width of $D_0(2400)^0$ ($\Gamma\sim 267$ MeV), which basically satisfies the necessary condition of forming a molecular state. Considering the above reasons, $D_0(2400)^0$ and $n$ can interact with each other before $D_0(2400)^0$ decays into other final states. This is why we still expect to discuss the molecular configuration to $X_c(3250)^0$ although $D_0(2400)^0$ is a very broad meson.

Since the total width of a state is mainly determined by its dominant decay, studying the dominant decay mode of $X_c(3250)^0$ provides the important information of its total decay width. What is more important is that this study can be as a critical test of the $n D_0^*(2400)$ molecular state assignment to $X_c(3250)^0$.

As the $nD_0^*(2400)$ molecular state, $X_c(3250)^0$ decays into $D^{+,0}\pi^{-,0}n$. Here, $X_c(3250)^0$ first falls apart into $n$ and $D_{0}^*(2400)^0$. Then, decays $X_c(3250)^0\to D^{+,0}\pi^{-,0}n$ occur via the intermediate $D_{0}^*(2400)^0$ as shown in Eq.(\ref{decay1}). Since the branching ratio of $D_{0}^*(2400)^0\to D\pi$ is almost 100\%, $X_c(3250)^0\to D^{+,0}\pi^{-,0}n$ is the dominant mode. For this process,
the differential decay width is written as
\begin{eqnarray}
&&d\Gamma\textnormal{\Huge{$[$}}
\raisebox{-23pt}{\includegraphics[width=0.14%
\textwidth]{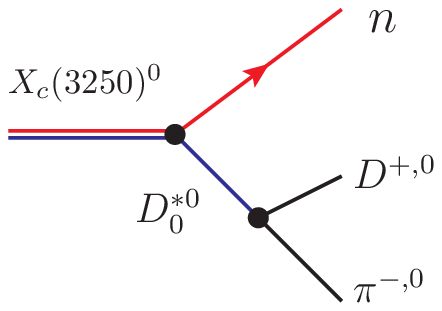}}
\textnormal{\Huge{$]$}}
={\frac{1}{2J+1}}\frac{1}{2E}(2\pi)^4{\sum_\lambda}|{\cal M}|^2\nonumber\\&&\qquad\times\delta^4\left(\sum_{i=1}^3q_i-P\right)
\frac{d^3q_1}{(2\pi)^32e_1}\frac{d^3q_2}{(2\pi)^32e_2}
\frac{d^3q_3}{(2\pi)^32e_3},
\label{decay1}
\end{eqnarray}
with $J=1/2$, where $q_{i} (e_{i})$ ($i=1,2,3$) denotes the momentum (energy) of final states. The decay amplitude $\mathcal{M}$ is expressed as
\begin{eqnarray}
{\cal M}&=&\frac{\mathcal{A}({X_c(3250)^0\to D_{0}^*n}) \,{\mathcal{A}}(D^*_0\to D\pi)}{q^2-M_{D^*_0}^2+iM_{D^*_0}\Gamma_{D^*_0}},\label{width}
\end{eqnarray}
where $\mathcal{A}_{X_c(3250)}\equiv \mathcal{A}({X_c(3250)^0\to D_{0}^*n})$ and $\mathcal{A}_{D^*_0}\equiv {\mathcal{A}}(D^*_0\to D\pi)$ describe the interactions $X_c(3250)^0\to n D^*_0(2400)^0$ and $D^*_0(2400)^0\to D^{+,0}\pi^{-,0}$, respectively. $M_{D^*_0}$ and $\Gamma_{D^*_0}$ denote the mass and width of the charmed meson $D_0^*(2400)^0$, respectively. $q$ is the four momentum carried by the off-shell $D_0^*(2400)^0$.

\begin{figure}[h!]
\begin{center}
\includegraphics[bb=60 625 750 785,scale=0.39,clip]{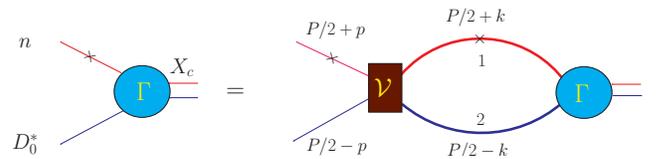}
\end{center}
\caption{(Color online.) Diagrammatic representation of the Gross equation for
the vertex function $\Gamma$. Here, $n$ and $D_0^*(2400)^0$ are marked by indexes 1 and 2, respectively.
$\times$ denotes $n$ being on-shell.
 \label{gross}}
\end{figure}

By the convariant spectator theory (CST), we can describe the collapse of $X_c(3250)^0$ into the on-shell $n$ and the $D_0^*(2400)$, where the vertex function $|\Gamma\rangle$ satisfies the relation
\begin{eqnarray}
|\Gamma\rangle=V G|\Gamma\rangle,
\end{eqnarray}
which is obtained by the Gross equation. In Fig. \ref{gross}, we present the diagrammatic representation of the Gross equation for the vertex function $\Gamma$.
We further obtain the Gross equation, which is dependent on three-momentum $\bm p$ as
\begin{eqnarray}
\Gamma ({\bm p})=\int\frac{d^3
	k}{(2\pi)^3} {\mathcal V}({\bm p},{\bm k},W)
	G({\bm k},W)~\Gamma({\bm k})
	\label{Eq: Gamma}
\end{eqnarray}
{with $\Gamma({\bm p})=\tilde{\Gamma}({\bm p}) u_R(W)$. We use the convention $\bar{u}u=2m$ here and hereafter,} where we perform the integration over $k^0$. $P=(W,\bm 0)$ denotes the four-momentum of the $n D_0^*(2400)^0$ system. $p=(p_1-p_2)/2$ and $k=(k_1-k_2)/2$ are the relative momenta as depicted in Fig. \ref{gross}.
Just because neutron is on the mass shell, one gets $k=(k_0,{\bm k})$, $p=(p_0,{\bm p})$, $k_0=E_1(k)-\frac{1}{2}W$, $E_1(k)=\sqrt{m_1^2+k^2}$. ${\mathcal V}({\bm p},{\bm k},W)$ is the interaction kernel with neutron on the mass shell.
The two-body Green's function $G({\bm k},W)$ in Eq.~(\ref{Eq: Gamma})
is expressed as
\begin{eqnarray}
	G({\bm k},W)&=&{\frac{1}{2E_{1}({\bm k}) }}u^{(t)}({\bm
	k})\bar{u}^{(t)}({\bm k})~\frac{1}{2E_{2}({\bm k})}
	\left(\frac{1}{E_{2}({\bm k})-E_{1}({\bm k})+W}\right.\nonumber\\&&\left.+\frac{1}{E_{2}({\bm k})+E_{1}({\bm k})-W}\right).\quad\label{Eq: Green}
\end{eqnarray}
The normalization of the vertex $\Gamma({\bm p})$ requires
\cite{He:2011ed,Stadler:1996ut},
\begin{eqnarray}
	1=\int\frac{d^3p}{(2\pi)^3}\Gamma^\dag({\bm p})
	\frac{\partial}{\partial W^2}[G({\bm p},W)]
	\Gamma({\bm p}).\label{n}
\end{eqnarray}

The wave functions for the $n D_0^*(2400)^0$ bound state are
\begin{eqnarray}
\psi^+_r({\bm
p})&=&{\frac{1}{\sqrt{(2\pi)^3 2W}}\sqrt{\frac{1}{2E_{1}({\bm p})2E_{2}({\bm p})}}
\frac{\bar{u}^{(r)}({\bm p})\tilde{\Gamma}({\bm p})u_R(W)}{ E_{2}({\bm p})+E_{1}({\bm p})-W}},\nonumber\\\label{w1}
\\
\psi^-_r({\bm
p})&=&{\frac{1}{\sqrt{(2\pi)^3 2W}}\sqrt{\frac{1}{2E_{1}({\bm p})2E_{2}({\bm p})}}
\frac{\bar{u}^{(r)}({\bm p})\tilde{\Gamma}({\bm p})u_R(W)}{E_{2}({\bm p})-E_{1}({\bm p})+W}}.\nonumber\\
\end{eqnarray}

With the above preparation, the integral equations are written as
\begin{eqnarray*}
&&\left[E_{2}({\bm p})+E_{1}({\bm p})-W\right] \psi^+_r({\bm p})\nonumber\\
&&=-\int\frac{d^3k}{(2\pi)^3}
	[{V}_{rr'}({\bm p},{\bm k},W)\psi^+_{r'}({\bm k})
 	+{V}_{rr'}({\bm p},{\bm k},W)\psi^-_{r'}({\bm k})],\quad\\
&&\left[E_{2}({\bm p})-E_{1}({\bm p})+W \right]\psi^-_r({\bm p})\nonumber\\
&&=-\int\frac{d^3k}{(2\pi)^3}
	[{V}_{rr'}({\bm p},{\bm k},W)\psi^+_{r'}({\bm k})
 	+{V}_{rr'}({\bm p},{\bm k},W)\psi^-_{r'}({\bm k})],\quad
\end{eqnarray*}
where the expression of potential ${V}_{rr'}({\bm p},{\bm k},W)$ is
\begin{eqnarray}
{V}_{rr'}({\bm p},{\bm k},W)&=&-{\sqrt{\frac{1}{2E_1({\bm p})}}\sqrt{\frac{1}{2E_2({\bm p})}}
\sqrt{\frac{1}{2E_1({\bm k})}}\sqrt{\frac{1}{2E_2({\bm k})}}}
\nonumber\\&&\times	\bar{u}^{(r)}({\bm p}){\cal V}({\bm p},{\bm k},W)u^{(r')}({\bm k}).\label{la}
\end{eqnarray}

After taking the nonrelativized approximation \cite{Gross:1969rv,Gross:1972ye,Buck:1979ff} and the Fourier transformation, we get the integral equations in the coordinate space
\begin{eqnarray}
    -\left(\frac{\nabla^2}{\mu}+\epsilon\right)\psi^+(\bm r)&=&
-\left[V(\bm r)+V(\bm r)F(\bm r)V(\bm r)\right]\psi^+(\bm r), \label{h1}\\
    \psi^-(\bm r)&=&-\left[F(\bm r)V(\bm r)\right]\psi^+(\bm
    r),\label{h2}
\end{eqnarray}
where $F(\bm r)=[m_2-m_1+W+V(\bm r)]^{-1}$ and $E_{1}(\bm p)+E_{2}(\bm
p)-W\approx -\epsilon+\frac{ {\bm p}^2}{\mu}$ with the reduced mass
$\mu$ and the binding energy $\epsilon=W-m_1-m_2$.

For the loosely bound system discussed in this work, it is reasonable to assume $m_{1,2}\gg \langle V\rangle$. Then, we have
\begin{eqnarray}
-\left(\frac{\nabla^2}{\mu}+\epsilon\right){\psi}^+(\bm r)&=&-V(\bm r){\psi}^+(\bm r),\label{hh1}\\
   { \psi}^-(\bm r)&=&0,
\end{eqnarray}
where Eq. (\ref{hh1}) corresponds to the Schr\"odinger equation. We need to specify that $V({\bm r})$ is the total effective potential of the $nD_0^*(2400)$ molecular state shown in Eq. (\ref{total}). Thus, by solving the Schr\"odinger equation with the deduced effective potential for the $nD_0^*(2400)$ molecular state, we obtain ${\psi}^+({\bm r})$. By the Fourier transformation, we get the wave function ${\psi}^+({\bm p})$ in the momentum space
\begin{eqnarray}
{\psi}^+({\bm p})&=&\frac{1}{(2\pi)^{3/2}}\int d^3re^{-i{\bm p}\cdot{\bm
r}}{\psi}^+({\bm r}),
\end{eqnarray}
which satisfies the normalization condition $\int d^3p~|{{\psi}^+}(\bm p)|^2=1$ required by
Eq. (\ref{n}), where ${\bm p}$ denotes the relative momentum.

With the calculated ${{\psi}^+}({\bm p})$ and Eq. (\ref{w1}), we obtain the vertex $\Gamma({\bm p})$,
which directly corresponds to $\mathcal{A}_{X_c(3250)}$ in Eq. (\ref{width}). By the CST, the calculated wave function is related to the vertex of the $X_c(3250)^0$ collapse into $nD_0^*(2400)^0$.
Thus, we find the relation
\begin{eqnarray}
&&\frac{ \mathcal{A}_{X_c(3250)}}{q^2-M_{D^*_0}^2+iM_{D^*_0}\Gamma_{D^*_0}}
\nonumber\\&&= {\frac{\sqrt{2W(2\pi)^3}\sqrt{2E_{n}({\bm q}) ~2E_{D^*_0(2400)}({\bm q})}
~\psi^+_r({\bm q})}
{-W+E_n({\bm q})-E_{D^*_0(2400)}({\bm q})}}.
\end{eqnarray}

In addition, the amplitude $\mathcal{A}_{D_0^{*}}$ reads as
\begin{eqnarray}
\mathcal{A}_{D_{0}^{*0}(2400)}&=&i\, I\,g_\pi\sqrt{M_{D_{0}^{*0}}M_{D}},
\end{eqnarray}
where the isospin factor $I$ is taken as 1 and $1/\sqrt{2}$ for decays $D_0^*(2400)\to D^+\pi^-$ and $D_0^*(2400)\to D^0\pi^0$, respectively. The coupling constant $g_\pi$ is determined by the decay width ($\Gamma_{D_0^*}=267\pm40$~MeV~\cite{Nakamura:2010zzi}), {\it i.e.},
\begin{eqnarray}
\Gamma(D_{0}^{*0}(2400)^0\rightarrow
D^{+,0}\pi^{-,0})
&=&\frac{M_{D^+}|{\bm p}|}{8\pi M_{D_{0}^{*}}}\,g_\pi^2
\end{eqnarray}
with ${\bm p}$ being the three-momentum of the daughter meson in the rest frame of the $D_{0}^{*0}(2400)$ meson, where the $D\pi$ channel contributes most to the total width of $D^*_0(2400)$ \cite{Close:2005se,Godfrey:2005ww}.

\begin{table}[htbp]
\caption{The decay width of $X_c(3250)^0\to n D\pi$ dependent on $\Lambda$ and the comparison of our result with the BaBar result. Here, we also calculate the binding energy $E$ and the decay width $\Gamma$ of $Y_c(3250)^+$, which is the isoscalar $D_0^*(2400)N$ molecular state defined in Eq. (\ref{s}) as the partner of $X_c(3250)$. $\Lambda$, $E$ and $\Gamma$ are in units of GeV, MeV and MeV, respectively. The effective potential of $Y_c(3250)^+$ can be easily obtained by replacing the factor in front of $\mathcal{V}_\rho(r)$ in Eq. (\ref{total}), {\it i.e.},  $1/2\to -3/2$. 
\label{Tab: decay width}}
\renewcommand\tabcolsep{0.3cm} \renewcommand{\arraystretch}{1.2}
\begin{center}
    \begin{tabular}{ ccc|ccc}  \toprule[1pt]
    \multicolumn{3}{c}{$X_c(3250)^0$ with $I=1$}&\multicolumn{3}{c}{$Y_c(3250)^+$ with $I=0$}\\\midrule[1pt]
     $\Lambda$& $E$& $\Gamma$&$\Lambda$& &$E$
     \\\midrule[1pt]
      1.17 & -3 &$121\pm18$ & 3.30&& -3
      \\
      1.20 & -7 &$111\pm17$ & 3.60&& -6
      \\
      1.23 & -13&$105\pm16$ & 4.20&&-14
      \\
      1.26 & -22&$100\pm15$ & 4.80&&-23
      \\
      1.30 & -35&$92\pm14$ & 5.70&&-35
      \\\hline
      BaBar \cite{babar}  & -13& $108\pm\ 6$ &- && -
      \\
\bottomrule[1pt]
\end{tabular}
\end{center}
\end{table}

In Table \ref{Tab: decay width}, for several typical values of $\Lambda$ we give the decay width of $X_c(3250)^0\to nD\pi$, which is the dominant decay mode of $X_c(3250)^0$ but does not strongly dependent on $\Lambda$. We compare the theoretical value of the decay width $X_c(3250)^0\to nD\pi$ with the BaBar's data \cite{babar}, which shows that the total width of $X_c(3250)^0$ under the assignment of the $nD_0^*(2400)^0$ molecular state is comparable with the BaBar's measurement \cite{babar}. The study of the dominant decay channels of $X_c(3250)^0$ also supports the $nD_0^*(2400)^0$ molecular state explanation for the observed $X_c(3250)^0$.

\begin{figure}[h!]
\begin{center}
\begin{tabular}{cc}
\includegraphics[bb=65 12 410 280,scale=0.37,clip]{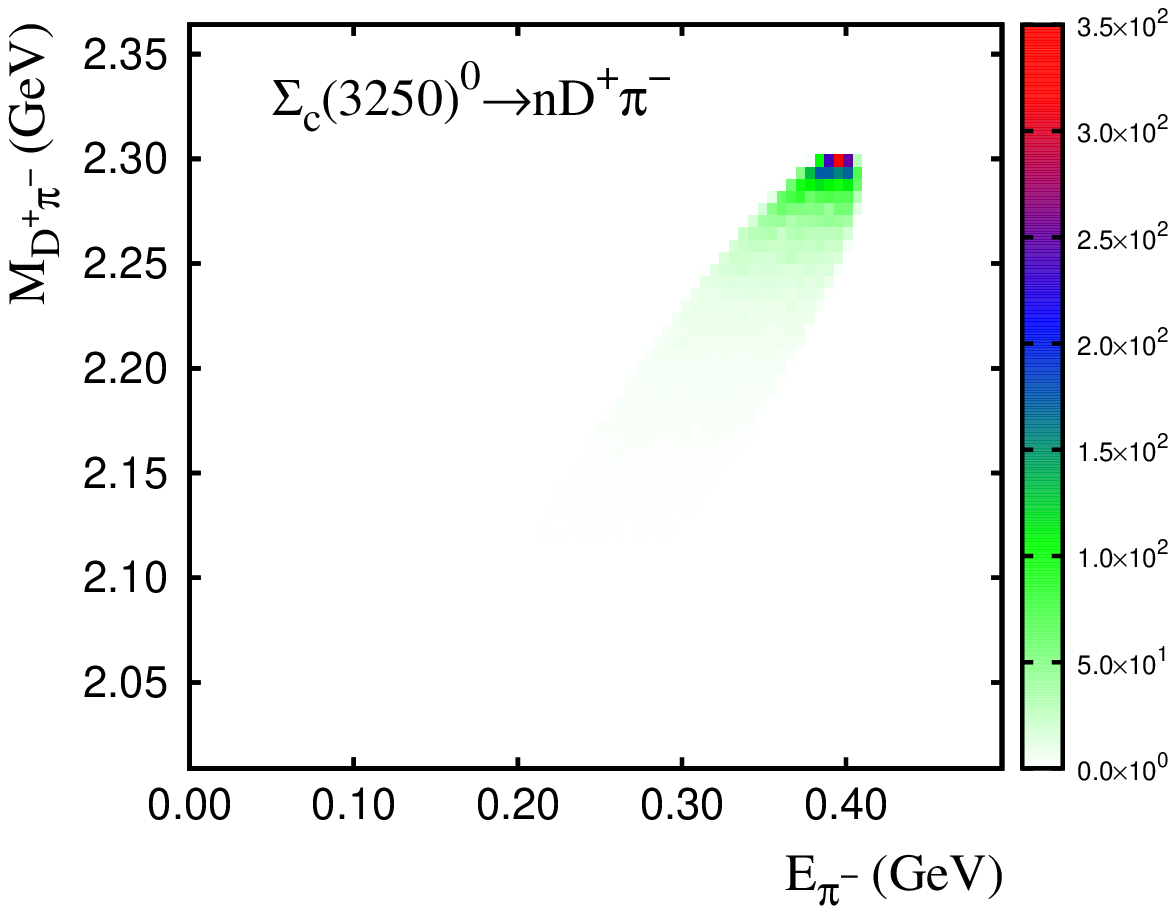}&
\includegraphics[bb=65 12 410 280,scale=0.37,clip]{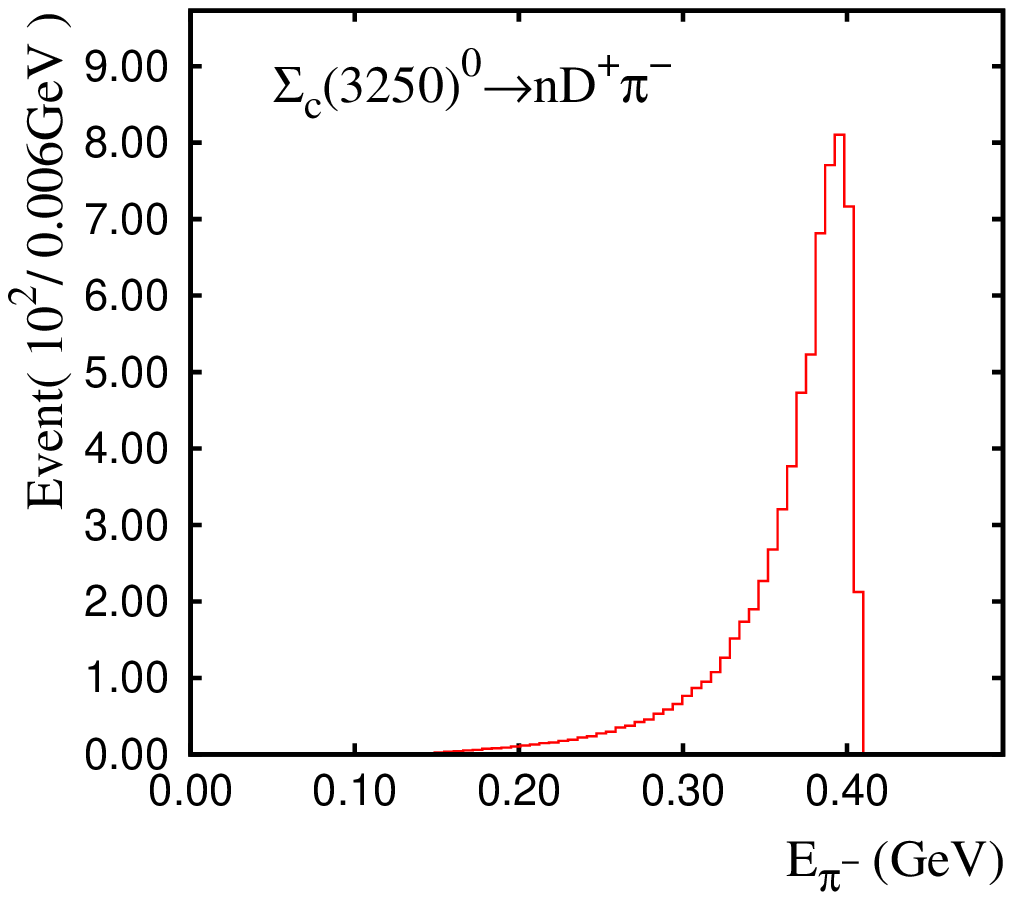}
\end{tabular}
\end{center}
\caption{(color online). The Dalitz plot analysis and the
pion spectrum for the $X_c(3250)^0\to nD^{+}\pi^-$ decay.  \label{decay}}
\end{figure}

Apart from calculating the dominant decay width of $X_c(3250)^0$, we also calculate the corresponding line shape of the pion spectrum of the $X_c(3250)^0\to nD^{+}\pi^-$ process with the help of the CERNLIB program FOWL (see Fig. \ref{decay}).

\begin{figure}[h!]
\begin{center}
\begin{tabular}{cccc}
\scalebox{0.36}{\includegraphics{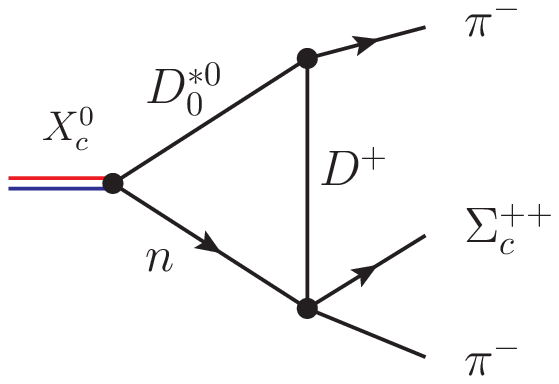}}&
\scalebox{0.36}{\includegraphics{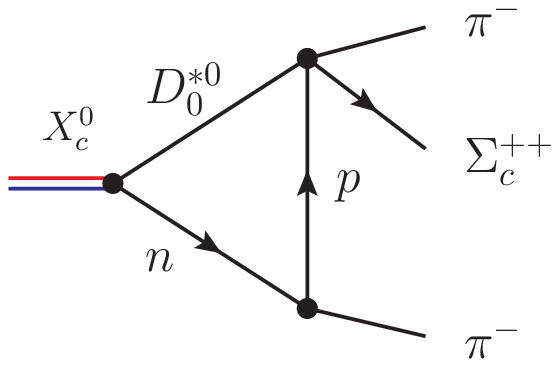}}&
\scalebox{0.36}{\includegraphics{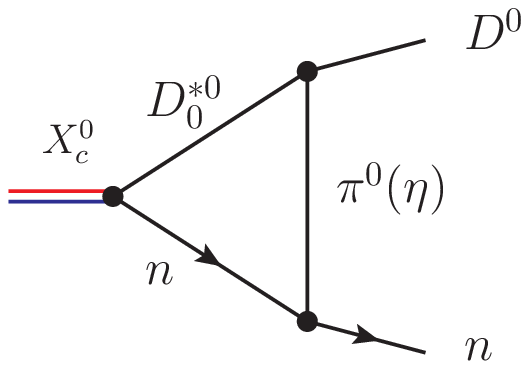}}&
\scalebox{0.36}{\includegraphics{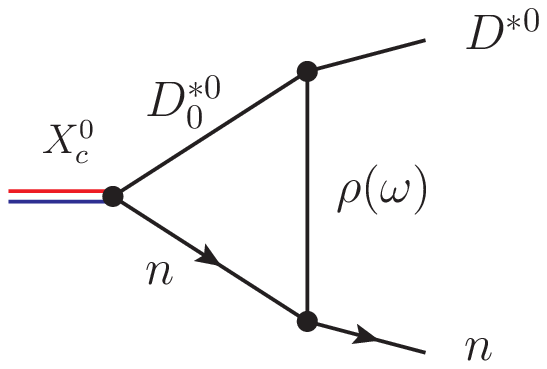}}
\\(a)&(b)&(c)&(d)\\
\raisebox{-1ex}{\scalebox{0.36}{\includegraphics{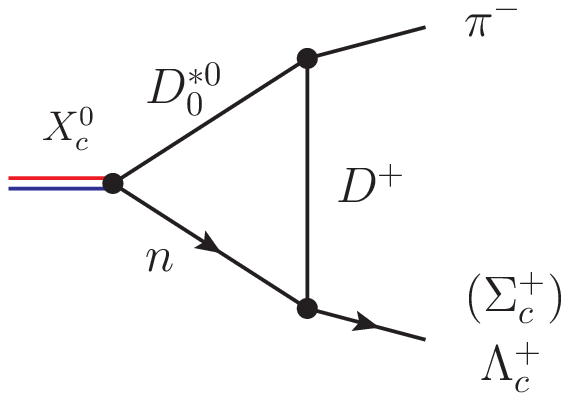}}}&\scalebox{0.36}{\includegraphics{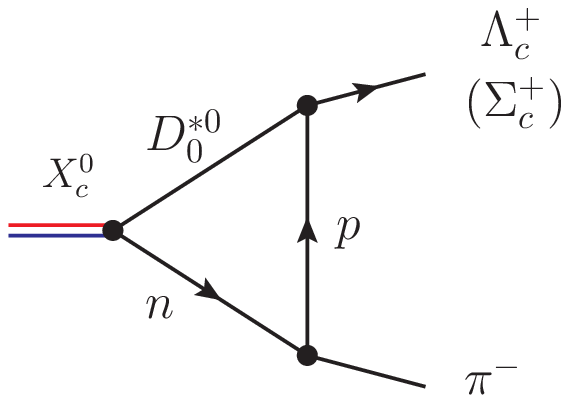}}&
\raisebox{-1ex}{\scalebox{0.36}{\includegraphics{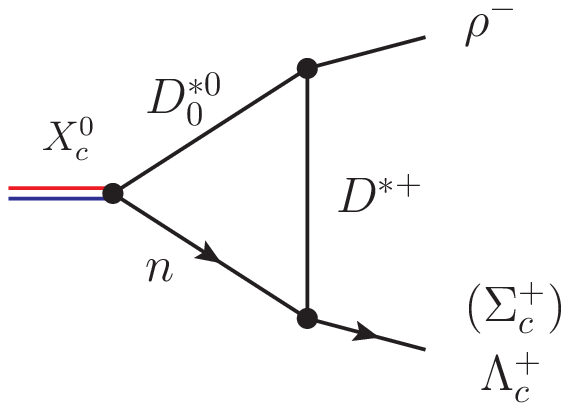}}}&
\scalebox{0.36}{\includegraphics{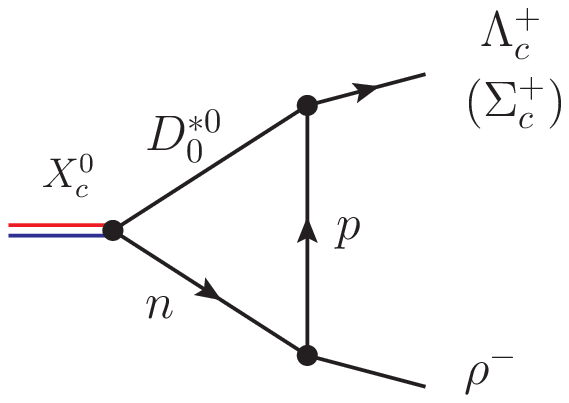}}
\\(e)&(f) & (g)&(h)
\end{tabular}
\end{center}
\caption{(color online). The $X_c(3250)^0\to\Sigma_c^{++}\pi^-\pi^-$ decay and other possible decays of $X_c(3250)^0$.  \label{od}}
\end{figure}

Since $X_c(3250)^0$ was observed in the $\Sigma_c^{++}\pi^-\pi^-$ invariant mass spectrum, in the following analysis we illustrate how this process can happen under the $nD_0^*(2400)^0$ molecular state assignment. Due to this molecular picture, $X_c(3250)^0 $ first disassociate into off-shell $n$ and $D_0^*(2400)^0$, which then transits into the final states $\Sigma_c^{++}\pi^-\pi^-$ by exchanging the $D^+$ meson (see Fig. \ref{od} (a)-(b) for more details). Thus, we can naturally explain why $X_c(3250)^0 $ is reported in the $\Sigma_c^{++}\pi^-\pi^-$ invariant mass spectrum.

We also qualitatively discuss other possible decay modes of $X_c(3250)^0$, which are shown in Fig. \ref{od} with the corresponding schematic diagrams. Here, the $X_c(3250)^0$ decays into $D^{0}n$, $D^{*0}n$, $\Lambda_c^+\pi^-$, $\Lambda_c^+\rho^-$, $\Sigma_c^+\pi^-$, $\Sigma_c^+\rho^-$ occur via these triangle diagrams listed in Fig. \ref{od} (c)-(h). We notice that these neutrons in the final states of the $X_c(3250)^0\to D^{0}n, D^{*0}n$ decays are neutral. Thus experimental search for these two decay modes of $X_c(3250)^0$ is difficult. Besides decaying into $D^{(*)0}n$, $X_c(3250)^0$ can decay into a charmed baryon plus a light meson such that pion and $\rho$. Searching for $X_c(3250)^0$ by these predicted decay channels will be an interesting research topic. Due to the subsequent decays $\Sigma_c\to \Lambda_c+\pi$ and $\rho\to 2\pi$, finally the processes of $X_c(3250)^0$ into $\Lambda_c^+\rho^-$, $\Sigma_c^+\pi^-$, $\Sigma_c^+\rho^-$
are related to these final states of a $\Lambda_c$ baryon plus multipion. Comparing with these processes, 
$X_c(3250)^0\to \Lambda_c^+\pi^-$ is a typical two-body decay. We suggest experiment to carry out the search for $X_c(3250)^0$ by its $\Lambda_c^+\pi^-$ invariant mass spectrum.

If $X_c(3250)^0$ is a $D_0(2400)^0 n$ molecular state, its dominant decay mode is $D\pi n$. In addition, we also listed its other decay modes (see Fig. \ref{od}), which occur via hadronic loop. Thus, these decay modes are suppressed compared with its $D\pi n$ decay. Although $D\pi n$ is its dominant decay mode, the final states of $X(3250)\to D\pi n$ contain neutral neutron, which makes that it is difficult to find its $D\pi n$ decay mode in experiment. It is the reason why experimental firstly reported $X(3250)$ in $\Sigma_c^{++} \pi^- \pi^-$.

In this work we also study the possible existence of the isoscalar partner of $X_{c}(3250)$, which is named as $Y_c(3250)^+$ whose flavor wave function is given in Eq. (\ref{s}). We find that the obtained $\Lambda$ value corresponding to its bound
state solution is around 3.3 GeV, which is larger than 1 GeV. Thus, according to the criteria ($\Lambda\sim 1$ GeV), we can conclude that it is impossible to form $Y_c(3250)^+$ molecular state.

Now let us draw a brief conclusion. Being stimulated by the recent observation of an enhancement structure $X_c(3250)^0$ in the $\Sigma_c^{++}\pi^-\pi^-$ invariant mass spectrum of $B^-\to \Sigma_c^{++}\bar{p}\pi^-\pi^-$ \cite{babar}, we find that $X_c(3250)^0$ can be well explained as the isovector $nD^{*}_0(2400)^0$ molecular hadron, which is supported by both the analysis of the mass spectrum and the study of its dominant decay channel. Furthermore, the observed $X_c(3250)^0\to \Sigma_c^++\pi^-\pi^-$ can be described reasonably under this picture. In addition, we also mention several other possible decay modes of $X_c(3250)^0$, which can be studied in future experiments.
We expect the contributions from BaBar, Belle, LHCb, and forthcoming BelleII, SuperB, which are the ideal places to further investigate the observed $X_c(3250)^0$ in the $B$ decay.

\section*{Acknowledgement}
We would like to thank Dr. Takayuki Matsuki for reading our manuscript. This project is supported by the National Natural Science Foundation of
China under Grants 11175073, 11035006, 10905077, 11005129, the
Ministry of Education of China (FANEDD under Grant No. 200924,
DPFIHE under Grant No. 20090211120029, NCET, the Fundamental
Research Funds for the Central Universities), the Fok Ying-Tong Education Foundation (No. 131006) and Chinese
Academy of Sciences (the West Doctoral Project and No. YZ080425).

\end{document}